\documentclass[aps,prl,twocolumn,superscriptaddress]{revtex4-1}

\usepackage{graphicx}
\usepackage{color}
\usepackage[version=4]{mhchem} 
\usepackage[caption=false]{subfig} 

\begin{document}




\title{Quantum Detection using Magnetic Avalanches in Single-Molecule Magnets}


\author{Hao Chen}
\author{Rupak Mahapatra}
\author{Glenn Agnolet}
\affiliation{Department of Physics and Astronomy, Texas A\&M University, College Station, TX, 77843}
\author{Michael Nippe}
\affiliation{Department of Chemistry, Texas A\&M University, College Station, TX, 77843}
\author{Minjie Lu}
\affiliation{Department of Physics and Astronomy, Texas A\&M University, College Station, TX, 77843}
\author{Philip C. Bunting}
\affiliation{Department of Chemistry and Biochemistry, University of California San Diego, La Jolla, California 92093}
\author{Tom Melia}
\affiliation{Kavli IPMU (WPI), UTIAS, The University of Tokyo, Kashiwa, Chiba 277-8583, Japan} 
\author{Surjeet Rajendran}
\affiliation{Department of Physics \& Astronomy, The Johns Hopkins University, Baltimore, Maryland 21218}
\author{Giorgio Gratta}
\affiliation{Physics Department, Stanford University, Stanford, CA, 94305}
\author{Jeffrey R. Long}
\affiliation{Department of Chemistry and Department of Chemical and Biomolecular Engineering, University of California, Berkeley, CA 94720 and Materials Sciences Division, Lawrence Berkeley National Laboratory, Berkeley, CA 94720.}


\date{\today}

\begin{abstract}

The detection of a single quantum of energy with high efficiency and low false positive rate is of considerable scientific interest, from serving as single quantum sensors of optical and infra-red photons to enabling the direct detection of low-mass dark matter. We report the first experimental demonstration of magnetic avalanches induced by scattering of quanta in single-molecule magnet (SMM) crystals made of Mn12-acetate, establishing the use of SMMs as particle detectors for the first time. While the current setup has an energy threshold in the MeV regime, our results motivate the exploration of a wide variety of SMMs whose properties could allow for detection of sub-eV energy depositions. 

\end{abstract}


\maketitle

It is scientifically challenging to develop sensors that can detect energy depositions as low as $\sim 10$~meV with  high efficiency and low false positive (or dark count) rates. Sensors with this capability can be used to count single quanta of infra-red photons, a technical feat that has broad applications to many fields \cite{Komiyama2000, Lita:s, Eisaman2011}, including quantum computing \cite{Ladd2010, Knill2001}. Such single-quantum sensors \cite{HVeV2019, Rong2018} may also open a path towards the detection of the scattering or absorption of low mass (sub GeV) dark matter particles. This is a theoretically well motivated region of dark matter parameter space that has so far not been well explored \cite{PhysRevLett.121.101801, Battaglieri:2017aum, PhysRevD.93.103520}. The detection of small energies can be accomplished through the use of an amplification technique that magnifies the effect of the initial energy deposition\cite{Chesi2019, Knopfmacher2014, Sorgenfrei2011}. 
Recently, it was proposed  that a high-gain, low-threshold detector that can detect energies as low as 10 meV but with a low false positive rate could be realized in single crystals of single-molecule magnets (SMMs) \cite{Bunting2017}.

\begin{figure}[b]
	\centering
	\includegraphics[width=8.6cm]{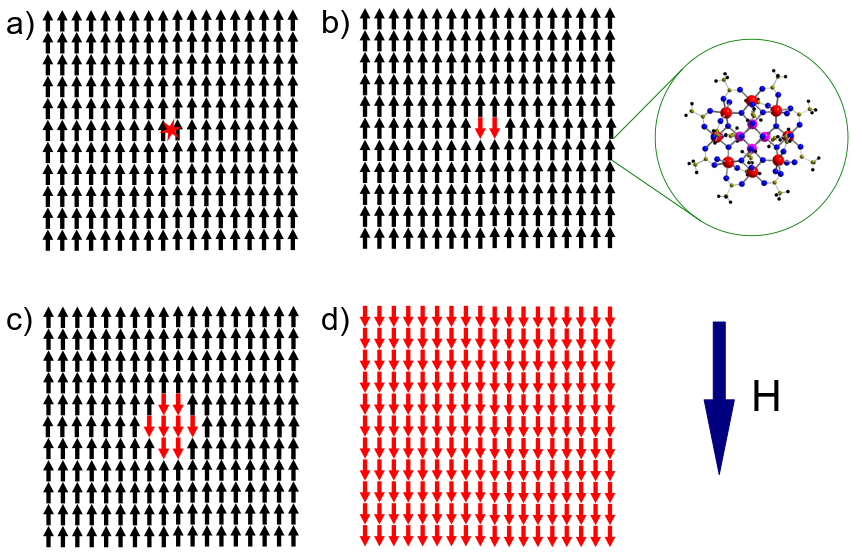}
	\caption{{\bf Conceptual illustration of a single-molecule magnet-based particle detector}. A crystal of the single-molecule magnets is first polarized at high temperature (3 K for Mn$_{12}$-acetate) and the magnetic field is then reversed after cooling to cryogenic temperatures. a) an interaction deposits some energy at a crystal site, b) the deposited energy locally heats the crystal, causing some of the spins to relax, releasing their Zeeman energy, c) the released energy further heats the crystal locally, causing nearby spins to also relax, d) The avalanche process continues until the whole crystal relaxes, with a measurable change in the crystal magnetization.}
	\vspace{10pt}
	\label{fig:avalanche}
\end{figure}

First discovered nearly 30 years ago~\cite{1993Natur.365..141S}, these unique compounds exhibit magnetic bistability and a barrier to magnetization reorientation, which can lead to phenomena such as magnetic hysteresis at low temperatures. Application of a static magnetic field lifts the degeneracy of the molecular magnetic ground state, giving rise to a metastable state that can persist for several months at cryogenic temperatures. As such, these molecules have garnered substantial interest for potential applications including spin-based electronics~\cite{Bogani2008} and quantum computing~\cite{Leuenberger2001}. While in this metastable state, certain single-molecule magnets can also undergo a rapid and complete reversal of their magnetization, resulting in the release of their Zeeman energy in a process known as a magnetic avalanche\cite{PhysRevLett.95.147201, PhysRevB.76.054410, PhysRevB.81.064437, PhysRevB.90.134405}. Magnetic avalanches have been studied in detail in the archetypal single-molecule magnet Mn$_{12}$O$_{12}$(O$_2$CCH$_3$)$_{16}$(H$_2$O)${_4}$ (Mn$_{12}$-acetate, Figure \ref{fig:avalanche})~\cite{1993Natur.365..141S}, triggered by mechanisms such as supplying energy via surface acoustic waves~\cite{PhysRevLett.95.217205}, direct heating of one side of a crystal~\cite{PhysRevLett.110.207203}, or sweeping the external magnetic field to directly alter the stability of the metastable state~\cite{PhysRevLett.95.147201}. In principle, a magnetic avalanche could also be triggered by small, localized energy depositions as recently proposed~\cite{Bunting2017} (Figure \ref{fig:avalanche}), enabling the use of single-molecule magnets as detectors for impinging particles or radiation in a manner analogous to superheated bubble chamber particle detectors~\cite{PhysRevD.100.022001}.

Herein, we report the first experimental demonstration~\cite{Chenthesis} of a magnetic avalanche triggered by $\alpha$ particle scattering in crystals of Mn$_{12}$-acetate. While the energy threshold of our particular setup is on the order of MeV, our results offer the first experimental proof-of-concept for the use of single-molecule magnets as single quantum sensors. These molecular magnets are set apart from other candidate sensors in the literature as a result of their unparalleled chemical tunability, and thus they have the potential to afford access to a versatile, tunable platform for next-generation quantum sensors, including for the detection of dark matter. 

\begin{figure}
	\centering
	\includegraphics[width=8.6cm]{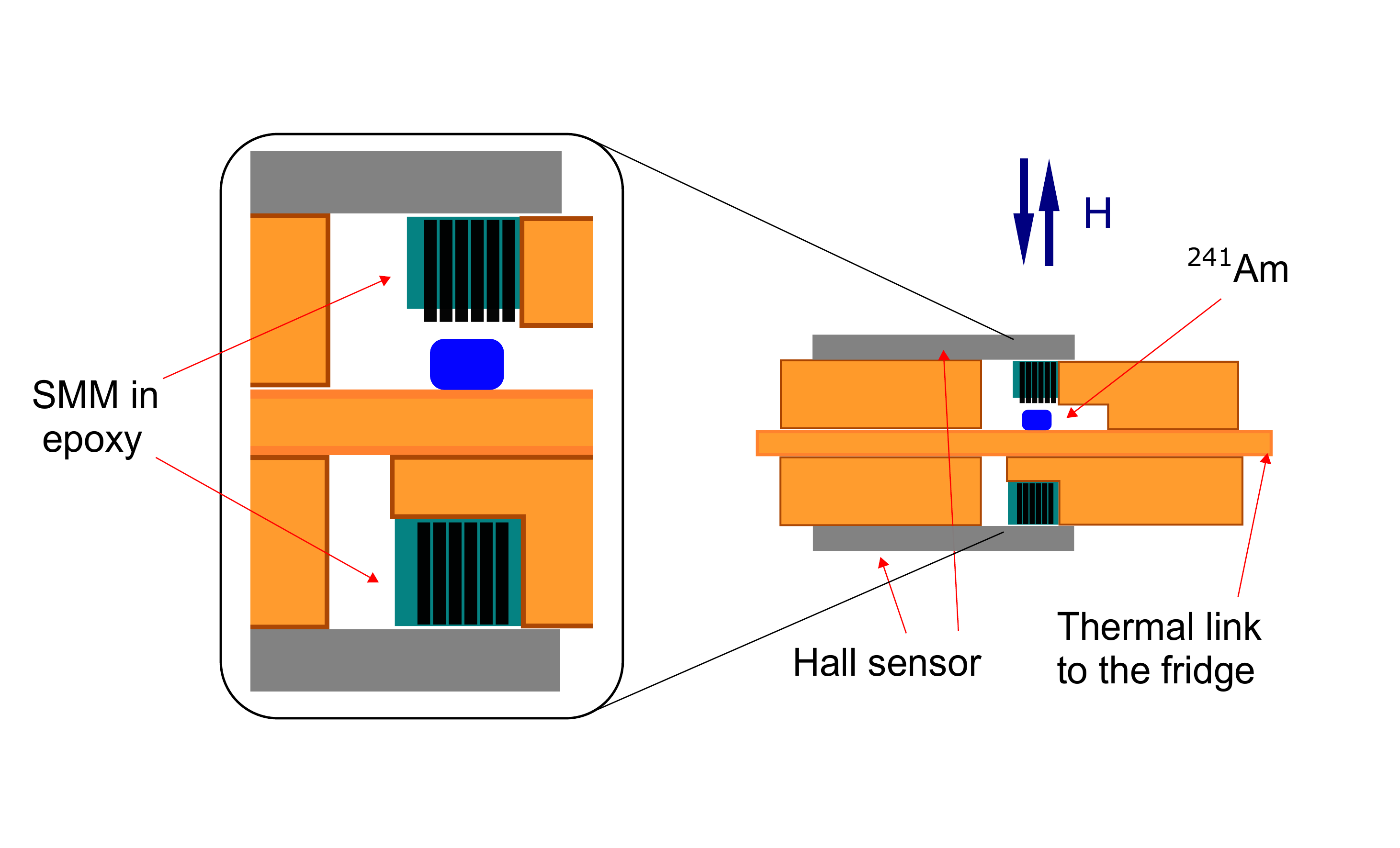}
	\caption{{\bf Schematic view of the experimental setup.} Crystals of Mn$_{12}$-acetate were mounted onto two sample holders connected to the same heat sink, each equipped with an independent Hall sensor. An $^{241}$Am $\alpha$ source was situated below the upper sample holder (Source), directly facing the crystals within. The crystals inside the lower sample holder (Control) were fully covered by epoxy and shielded from the source.}
	\label{fig:diagram}
\end{figure}

The particle detector setup, illustrated in Figure \ref{fig:diagram}, features two $3\times3\times3$ mm$^3$ crystal sample holders and thermal links made from oxygen-free, high-conductivity copper to ensure a diamagnetic background and optimal heat conduction. The crystal holders are mounted symmetrically and each has its own Hall sensor, so that particles can be detected in one crystal sample while the other is used as a control. Each Hall sensor is in close proximity ($\sim$ mm) to the exposed faces of the crystals to maximize the magnetic signal sensitivity, while thermally isolating it to ensure that Joule heating will not warm the crystals. Each holder can accommodate a dozen Mn$_{12}$-acetate crystals ($\sim$2 mm long and 0.5-1 mm wide, as prepared~\cite{Lis:a19066}), which are held in place and connected to a heat sink using epoxy resin. Importantly, the molecules crystallize such that the long dimension is aligned with the molecular easy axis -- that is, the axis along which the crystals are readily magnetized. In our experiments, the Mn$_{12}$-acetate crystals were physically aligned in the direction of the external magnetic field in order to maximize the signal. Cooling in this setup is afforded via a connection to a $^3$He cryostat, and all experiments were conducted with the sample holders at 1.8 K, as measured by a silicon diode thermometer mounted nearby. The cryostat is equipped with a superconducting magnet that can generate magnetic fields as high as 40 kG, and appropriate controls are available to scan the magnetic field at preset rates. Given the sizes of the crystals used in this study, the change in the magnetic field near the Hall sensors caused by a magnetic avalanche can be up to 200 G. The Hall sensors were tested to ensure good linearity in the field range of 0-30 kG and for noise levels as high as 20 G, sufficient for the purposes of this work.

The $^{241}$Am source used in this setup emits $\alpha$ particles with kinetic energy of 5.486 MeV. Care was taken to ensure that the epoxy in the upper sample (called source sample) did not cover the faces of the crystals that were directly exposed to the source. In the lower sample (called control sample), the crystals were shielded from the $\alpha$ particles by the copper plate separating the source and the bottom sample holder. Thus, particle-triggered avalanches were expected to occur only in the sample holder exposed to the source. At cryogenic temperatures, it was expected that the metastable state of Mn$_{12}$-acetate would be sufficiently long-lived for meaningful data collection (on the order of a few months~\cite{Bunting2017, Sessoli1993}). Additionally, the low activity $\alpha$ particle source is expected to produce a low enough rate of less than one per minute to leave substantial time to assess noise in the cryostat that is correlated in the two channels.

Prior to data collection, the crystals were initialized by first heating with a resistor and then the external magnetic field was ramped from 0 to +10 or -10 kG. The threshold to trigger an avalanche in Mn$_{12}$-acetate depends on the field and is difficult to estimate under our conditions, largely because the thermal conductivity of the crystals in the epoxy is unknown. As such, for subsequent data collection, the field was then either set to 0 G and continuously ramped to increasing reverse values or reversed and held at discrete values while the Hall probes were monitored for avalanches. In the latter scenario, the reverse field was set to the given value by ramping in small discrete steps, and then the sample was held at this value for 6 to 10 min. If no avalanche signal was observed, the reverse field was increased by 100 G and held again for 6 to 10 min, and this process was repeated until a signal was observed. A total of four continuous field scan experiments and seven discrete step experiments were carried out. Prior to each run, the crystals were initialized as described above. 

\begin{figure}

\centering
	\includegraphics[width=8.6cm]{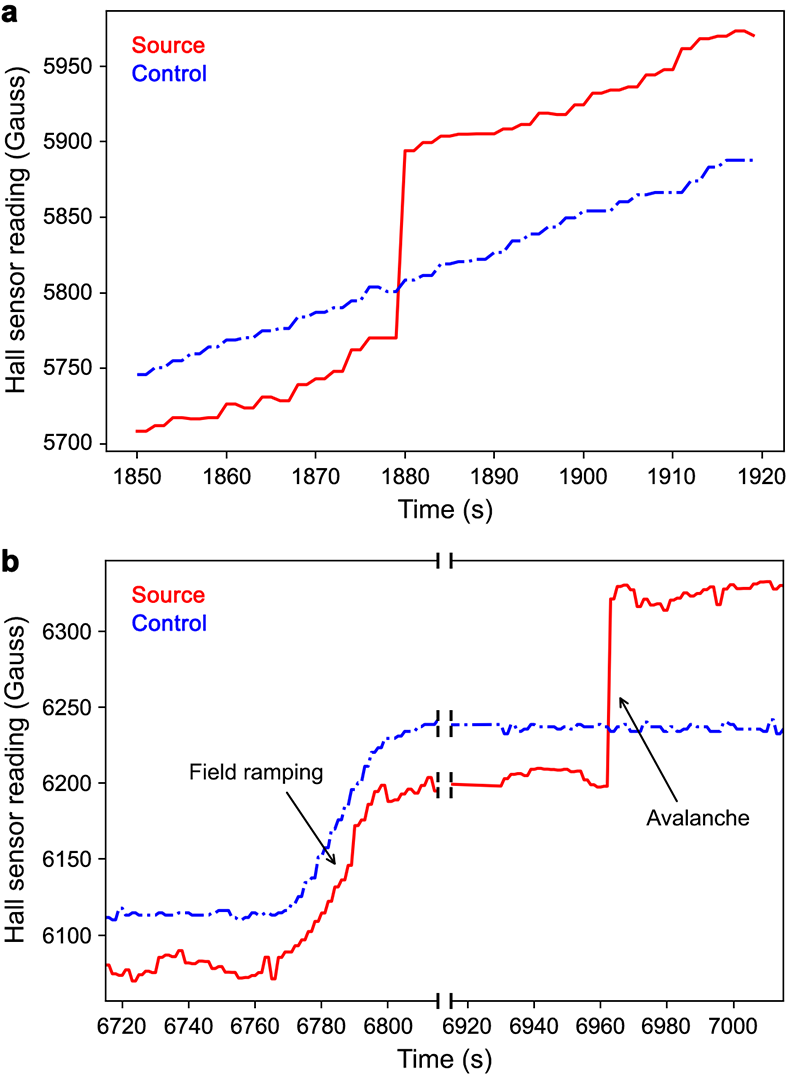}

\caption{\small {\bf Example of an $\alpha$ particle induced avalanche observed in \ce{Mn12}-acetate}. 
		a) Avalanche observed using a continuous reverse field ramp. The change in magnetization occurs in less than 1 s. There is no signal in the control sample at the time of the avalanche. The difference in the calibrated magnetic fields in the experimental and control Hall sensors is due to the slightly different locations of the samples in the cryostat. b) A particle-induced avalanche was also observed when the field was held at a constant value, after scanning the field in small discrete steps. Again, no signal was observed in the control sample at the time of the avalanche. The slow increase in the Hall probe signal in both Source and Control sample is due to the external field ramp up, whereas the sharp change is due to a magnetic avalanche that occurs in less than 1 s.}
\label{fig:scanfield}
\end{figure}

During each of the four continuous field scans (sweep rate of 2.1 G/s), an avalanche event was observed between 5100 and 6150 G in the crystals exposed to the $^{241}$Am source. In contrast, no avalanche was observed in the control sample.  Representative data from one of these experiments is shown in Figure~\ref{fig:scanfield}. The magnetic field jump for each avalanche ranged from 120 to 140 G for each scan, consistent with that expected given the size of the Mn$_{12}$-acetate crystals used here. 

Similar results were obtained when ramping the magnetic field in discrete steps and holding the field constant. Avalanches were observed in six of seven scans for the source sample—once while the field was held constant (see Figure~\ref{fig:scanfield}b) and on five other runs while ramping the field to a constant value. The maximum reversed field applied in the run without any observed avalanche was 5000 G, which might have prevented the detector from reaching the desired threshold for avalanche production. For these scans, the magnetic field threshold was ca. 4500--6300 G. The signal was observed clearly above the noise level. Again, no avalanche occurred in the control sample under these conditions.

In all 10 runs that resulted in an avalanche, this process occurred only once for the source sample in each run, indicating that all molecules in the sample underwent a spin flip each time an avalanche occurred. Also, in all cases the magnetic field jumps were abrupt (occurring in less than 1 s) and comparable to those observed in the literature\cite{PhysRevLett.95.147201, McHugh2007} for magnetic avalanches triggered in single-molecule magnets by other means. Moreover, given the stability of the control sample, we can be confident that other magnetic relaxation pathways, such as resonant quantum tunneling, do not play a role at these temperatures.  We note that it was possible to trigger a field-induced avalanche in the control sample by sweeping the magnetic field to higher magnitudes following the avalanche in the source sample. All together, these data strongly support the presence of magnetic avalanches in the Mn$_{12}$-acetate crystals induced by the absorption of $\alpha$ particles.

\begin{figure}
	\begin{center}
		\includegraphics[width=8.6cm]{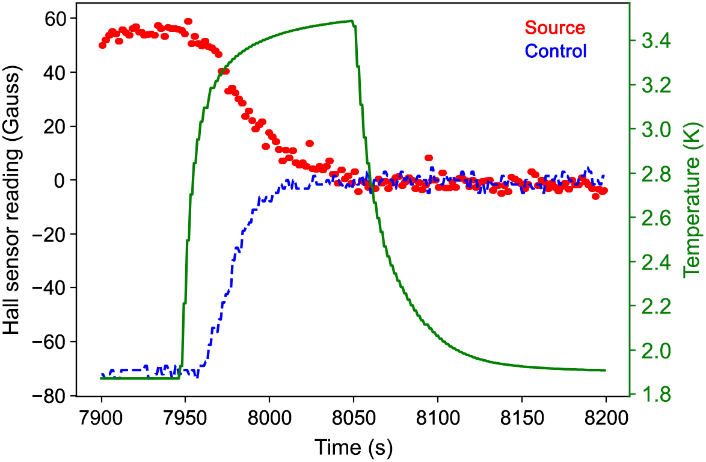}
		\caption{\small  {\bf Demagnetization of Mn$_{12}$-acetate crystals following observation of an avalanche in the source sample}. The sign for the source and the control sides are opposite, as expected. The amplitude of the transition for the crystals on the source side is 60 G, consistent with approximately half the value measured in an avalanche. The small difference in the absolute value of the demagnetization for the source and the control sample is consistent with the uncertainty in the size of the crystals and the relative positions of the Hall probes.}
		\label{fig:magentization}
	\end{center}
\end{figure}

To further support this conclusion, after each run, we measured the field change in the source and control sample while they were demagnetized with heating (see Figure~\ref{fig:magentization}). As expected, the samples exhibited opposite changes in field during demagnetization for the runs where only the source sample underwent an avalanche. This change is expected, given that both sets of crystals were initially magnetized by the same field, but only the source sample experienced spin reversal due to an $\alpha$ particle-induced avalanche. In contrast, for the two runs where a field-induced avalanche was also triggered in the Control, the samples exhibit like changes in field during demagnetization. We also note that the Hall sensor reading for the demagnetization (a change in the magnetization from $-\mathbf{M}$ to $0$) of the source crystal sample was $\sim$ 60 G (20--60 G for the ten runs), approximately half of the value of the full magnetization reversal in an avalanche ($+\mathbf{M}$ to $-\mathbf{M}$) as expected. Some runs had lower demagnetization values, which could be due to the crystals being partially demagnetized while the reverse field was quickly reduced to zero.
We note that the probability that the observed transitions are due to some random phenomenon common to the two samples (e.g., vibrations, electrical glitches, unstable temperature, or background radiation) can be estimated (using standard Poisson statistics) from the control channel and is negligible.

The foregoing results definitively show that magnetic avalanches in crystals of Mn$_{12}$-acetate can be triggered by the absorption of elementary particles, with a threshold lower than 5.486 MeV for $\alpha$ particles and the values of magnetic field and temperature reported here. We note that the experimental energy detection threshold determined here for Mn$_{12}$-acetate is high compared to the 10 meV threshold desired for single infra-red photon or dark matter ~\cite{JUNGMAN1996195}. 
There is considerable room for exploration in this area. The expected threshold for an avalanche in single-molecule magnets depends on many parameters, such as its molecular mass, thermal properties, and energy barrier.   
To develop lower threshold detectors, we will explore a large number of available single-molecule magnets  known to exhibit magnetic avalanches, optimizing on the magnet's energy barrier and thermal conductivity to achieve low thresholds while ensuring stability. 
Rigorous analysis will require determination of the specific heat capacities and thermal conductivities of candidate systems to better identify systems with appropriate threshold energies for detection of particles of various energies. 

 Given the metastability of the excited Zeeman state in single-molecule magnets at cryogenic temperatures, these molecules could act as high efficiency, low dark count single photon detectors for infrared photons. As dark matter detectors, they could search for the absorption of meV scale “dark photons” or the scattering of keV to MeV mass dark matter particles. Importantly, such molecular arrays should be sensitive to localized energy depositions. This property is important to ensure the suppression of backgrounds caused by electron scattering, wherein the same energy would be deposited in larger regions with a density that is insufficient to trigger the magnetic avalanche. Moreover, through the use of precision magnetometers, it may also be possible to identify the location of the scattering event in a bulk volume, so that surface events can be identified, enabling the demarcation of a low background fiducial volume in the bulk.

\begin{acknowledgments}
This work was supported by the Strategic Transformative Research Program at Texas A\&M University (R.N. and M.N.). R. M. acknowledges DOE support through awards DE-SC0017859 and DE-SC0018981 that were instrumental in providing equipment and facilities to carry out this experiment. M.N. is also supported by funding from the Welch foundation (A-1880). T.M. was supported by the World Premier International Research Center Initiative (WPI) MEXT, Japan, and by JSPS KAKENHI grants JP18K13533, JP19H05810, JP20H01896 and JP20H00153. S.R. was supported in part NSF grant PHY-1638509, the Simons Foundation (award 378243), and the Heising-Simons Foundation grants 2015038 and 2018-0765. P.C.B. and J.R.L. were supported under NSF grant CHE-1800252.
\end{acknowledgments}

\bibliography{mybibfile}

\end{document}